\documentclass[a4paper]{article}
\usepackage{RR}
\usepackage{hyperref}

\usepackage{multirow}
\usepackage{times}
\usepackage{epsfig}
\usepackage{./slashbox} 

\newcommand{\debliste}{\begin{list}{$-$}{\setlength{\parsep 0pt}{\itemsep 0pt}}} 
\newcommand{\finliste}{\end{list}}

\def\sqw{\hbox{\rlap{\leavevmode\raise.3ex\hbox{$\sqcap$}}$%
\sqcup$}}
\def\cqfd{\ifmmode\sqw\else{\ifhmode\unskip\fi\nobreak\hfil
\penalty50\hskip1em\null\nobreak\hfil\sqw
\parfillskip=0pt\finalhyphendemerits=0\endgraf}\fi}
\RRdate{Juin 2008}
\RRNo{6574}
\RRauthor{% les auteurs
Damien Hardy%Jos\'e Grimm\thanks{Footnote for first author}%
  \and
Isabelle Puaut %\thanks{Footnote for second author}%
}
\authorhead{Hardy \& Puaut}
\RRtitle{Analyse pire cas des hi\'erarchies de caches d'instruction associatifs par ensemble}
\RRetitle{WCET analysis of multi-level set-associative instruction caches}
\titlehead{WCET analysis of multi-level set-associative instruction caches}

\RRnote{This study was partially supported by the french
National Research Agency project Mascotte (ANR-05-PDIT-018-01)}
\RRresume{Avec l'arriv\'ee de mat\'eriel complexe dans les syst\`emes temps-r\'eel embarqu\'es
(processeurs avec des fonctions d'am\'elioration des performances tel que les pipelines, les hi\'erarchies
de caches, les multi-c\oe urs), de nombreux processeurs ont maintenant des caches L2 associatifs par ensemble.
Ainsi, consid\'erer les hi\'erarchies de caches lors de la validation du comportement temporel des
syst\`emes temps-r\'eel, en particulier lors de l'estimation d'une borne sup\'erieure du pire temps d'ex\'ecution des t\^aches
s'ex\'ecutant sur le syst\`eme devient n\'ecessaire. A notre connaissance, il existe une seule approche traitant des hi\'erarchies de caches
pour le calcul de cette borne~\cite{mueller97timing}, qui s'av\`ere \^etre non s\^ure pour les caches associatifs par ensemble.

Dans ce rapport, nous pr\'esentons les conditions pour lesquelles l'approche d\'ecrite dans~\cite{mueller97timing} est
non s\^ure. Une approche statique s\^ure est pr\'esent\'ee pour les caches d'instruction. A l'oppos\'e de~\cite{mueller97timing},
notre m\'ethode supporte les caches associatifs par ensemble et les caches totalement associatifs. Cette m\'ethode est exp\'eriment\'ee
sur des programmes de test ainsi qu'une application r\'eelle. Nous montrons que notre m\'ethode est la plupart du temps pr\'ecise
et l'estimation du pire temps d'ex\'ecution est toujours plus pr\'ecise en consid\'erant la hi\'erarchie de cache comparativement \`a un seul niveau de cache.
Une \'evaluation du temps de calcul est r\'ealis\'ee montrant que l'analyse de la hi\'erarchie de cache est effectu\'ee en un temps raisonnable.
}
\RRabstract{With the advent of increasingly complex hardware in real-time
  embedded systems (processors with performance enhancing features
  such as pipelines, cache hierarchy, multiple cores), many
  processors now have a set-associative L2 cache.  Thus, there is a need for considering cache hierarchies
  when validating the temporal behavior of real-time systems, in
  particular when estimating tasks' worst-case execution times
  (WCETs). To the best of our knowledge, there is only one approach
  for WCET estimation for systems with cache
  hierarchies~\cite{mueller97timing}, which turns out to be unsafe for set-associative caches.

  In this paper, we highlight the conditions under which the approach
  described in~\cite{mueller97timing} is unsafe. A safe static
  instruction cache analysis method is then presented. Contrary
  to~\cite{mueller97timing} our method supports set-associative and
  fully associative caches. The proposed method is experimented on
  medium-size and large programs. We show that the method is most of the time tight. 
  We further show that in all cases WCET estimations
  are much tighter when considering the cache hierarchy than when
  considering only the L1 cache. An evaluation of the analysis time is
  conducted, demonstrating that analysing the cache hierarchy has a reasonable computation time.}
\RRmotcle{pire temps d'ex\'ecution,, temps-r\'eel strict, hi\'erarchie m\'emoire, analyse statique, interpr\'etation abstraite.}
\RRkeyword{WCET, hard real time systems, memory hierarchy, static analysis, abstract interpretation.}
\RRprojet{CAPS}  % cas d'un seul projet
 \RRtheme{\THCom} % cas d'un seul theme
 \URRennes  % pour ceux qui sont \`a l'ouest
\begin{document}
\makeRR   % cas d'un rapport de recherche
%% \makeRT % cas d'un rapport technique.
%% a partir d'ici, chacun fait comme il le souhaite
\section{Introduction}
\label{sec:intro}

Cache memories have been introduced to decrease the access time to the 
information due to the increasing gap between fast micro-processors
and relatively slower main memories. Caches are very efficient at
reducing average-case memory latencies for applications with good
spatial and temporal locality. Architectures with caches are now
commonly used in embedded real-time systems due to the increasing
demand for computing power of many embedded applications.

In real-time systems it is crucial to prove that the execution of a
task meets its deadline in all execution situations, including the
worst-case. This proof needs an estimation of the worst-case execution
times (WCETs) of any sequential task in the system. WCET estimates have
to be safe (larger than or equal to any possible execution time). Moreover,
they have to be tight (as close as possible to the actual worst-case
execution time) to correctly dimension the ressources required by the
system.

The presence of caches in real-time systems makes the estimation of
both safe and tight WCET bounds difficult due to the dynamic behavior
of caches. Safely estimating WCET on architectures with caches
requires a knowledge of all possible cache contents in every execution
context, and requires some knowledge of the cache replacement policy.

During the last decade, many research has been undertaken to predict
WCET in architecture equipped with caches. Regarding instruction
caches, static cache analysis methods have been designed, based on
so-called \textit{static cache simulation}
\cite{WHIT:97a,mueller00timing} or \textit{abstract interpretation}
\cite{theiling00fast,FERD:01a}. Approaches for static data cache
analysis have also been proposed \cite{325479,1049905}. Other
approaches like cache locking have been suggested when the replacement
policy is hard to predict precisely \cite{PUAU:06a} or for data caches
\cite{VERA:03b}. The impact of multi-tasking has also been considered
by approaches aiming at statically determining cache related
preemption delays \cite{TUL:03,STA:05}.

To the best of our knowledge, only \cite{mueller97timing} deals with 
cache hierarchies. In this work, static cache analysis is applied to
every level of the cache hierarchy. The memory reference stream considered by the analysis at
level $L$ of the cache hierarchy (for example L2 cache) is a subset of
the memory reference stream considered at level $L-1$ (for example L1 cache) 
when the analysis ensures that some references always hit at level $L-1$.
However, we show that the way references are filtered out
in~\cite{mueller97timing} is unsafe for set-associative
caches. In this paper, we overcome this limitation through the
proposal of a safe multi-level cache analysis of the cache structure
for set-associative caches, whatever the degree of associativity. Our
approach can be applied to caches with different replacement policies 
thanks to the reuse of an existing cache analysis method.

The paper presents experimental results showing that in most ot the cases the
analysis is tight. Furthermore, in all cases WCET estimations are much
tighter when considering the cache hierarchy than when considering the
L1 cache only. 
An evaluation of the analysis time is also presented, demonstrating that analysing the L2 cache has a reasonable computation time.

The rest of the paper is organized as follows. Related work is
surveyed in Section~\ref{sec:related}.  Section~\ref{sec:limitation}
presents a counterexample showing that the approach presented
in~\cite{mueller97timing} may produce underestimated WCET estimates
when analysing set-associative caches. Section~\ref{sec:solution} then details our proposal.
Experimental results are given in Section~\ref{sec:perfs}. Finally,
Section~\ref{sec:conclu} concludes with a summary of the contributions
of this paper, and gives directions for future work.

%------------------------------------------------------------------------- 
\section{Related work}
\label{sec:related}

Caches in real-time systems raise timing
predictability issues due to their dynamic behavior and their
replacement policy. Many static analysis methods have been proposed in
order to produce a safe WCET estimate on architectures with
caches. 

To be safe, existing static cache analysis methods determine \textit{every}
possible cache contents at every point in the execution, considering
all execution paths altogether. Possible cache contents can be
represented as sets of \textit{concrete cache states} \cite{TUL:03} or by a more
compact representation called \textit{abstract cache states (ACS)}
\cite{theiling00fast,FERD:01a,mueller97timing,mueller00timing}.

Two main classes of approaches \cite{theiling00fast,mueller00timing} exist for static WCET analysis on architectures with caches.  

In \cite{theiling00fast} the approach is based on abstract
interpretation~\cite{CousotCousot77-1,CousotCousot04-WCC} and uses ACS. An $Update$
function is defined to represent a memory access to the cache and a
$Join$ function is defined to merge two different ACS in case
there is an uncertainty on the path to be followed at run-time
(e.g. at the end of a conditional construct).  In this approach, three
different analyses are applied which used fixpoint computation to determine:
if a memory block is \textit{always present} in the cache (\textit{Must} analysis), 
if a memory block \textit{may} be present in the cache (\textit{May} analysis), 
and if a memory block will not be evicted after it has been first loaded (\textit{Persistence}
analysis). A \textit{cache categorisation} (e.g. \textit{always-hit},
\textit{first-miss}) can then be assigned to every instruction based
on the results of the three analyses.  This approach originally designed for LRU caches has been extended for different cache replacement policies in \cite{HECK:03}: Pseudo-LRU, Pseudo-Round-Robin.  To our
knowledge, this approach has not been extended to analyze multiple levels
of caches.  Our multi-level cache analysis will be defined as
an extension of \cite{theiling00fast}, mainly because of the theoretical
results applicable when using abstract interpretation.

In \cite{mueller94static,mueller00timing}, so-called \textit{static
  cache simulation} is used to determine every possible content of the
cache before each instruction. Static cache simulation computes
abstract cache states using dataflow analysis.  A \textit{cache
  categorisation} (\textit{always-hit}, \textit{always-miss},
\textit{first-hit} and \textit{first-miss}) is used to classify the
worst-case behavior of the cache for a given instruction. The base
approach, initially designed for direct-mapped caches, was later 
extended to set-associative caches~\cite{WHIT:97a}.

The cache analysis method presented in \cite{mueller94static} has been
extended to cache hierarchies in \cite{mueller97timing}. A separate
analysis of each memory level is performed by first analysing the
behavior of the L1 cache. The result of the analysis of the L1 cache
is consequently used as an input to the analysis of L2 cache, and so
on. The approach considers an access to the next level of the memory
hierarchy (e.g. L2 cache) if the access is not classified as
\textit{always-hit} in the current level (e.g. L1 cache). As shown in
Section~\ref{sec:limitation}, this filtering of memory accesses,
although looking correct at the first glance, is unsafe for
set-associative caches. Our work is based on the same principles as
\cite{mueller97timing} (cache analysis for every level of the memory
hierarchy, filtering of memory accesses), except that the unsafe
behavior present in \cite{mueller97timing} is removed. Moreover, our
paper presents an extensive evaluation of the performance of
multi-level cache analysis, both in terms of tightness, and in terms
of analysis time.

\begin{figure*}[htbp]
\begin{center}
\psfig{width=0.8\textwidth,figure=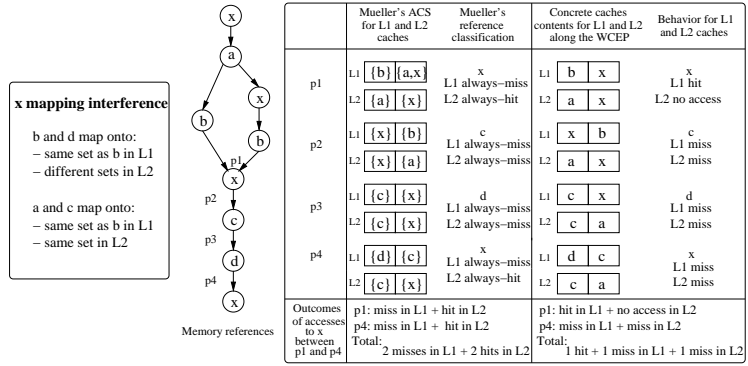}
\end{center}
\vspace*{-0.5cm}
\caption{Example of limitation for 2-ways L1 and L2 caches\label{fig:limit}}
\end{figure*}

%------------------------------------------------------------------------- 
\section{Limitation of Mueller's approach}
\label{sec:limitation}

The multi-level cache analysis method presented by F.~Mueller in
\cite{mueller97timing} performs a separate analysis for each level in
the memory hierarchy. The output of the analysis for level $L$ is a
classification of each memory references as \textit{first-miss},
\textit{first-hit}, \textit{always-miss}, or \textit{always-hit}, and is used
as an input for the analysis of level $L+1$. In \cite{mueller97timing} \textit{always-hit} means
that the reference is guaranteed to be in the cache;
\textit{always-miss} is used when a reference is not guaranteed to be
in the cache (but may be in the cache for some execution paths);
\textit{first-hit} and \textit{first-miss} are used for references
enclosed in loops, to distinguish the first execution from the others.
All references are considered when analyzing level $L+1$ exept those classified as \textit{always-hit} at level $L$ (or at a previous level). The implicit assumption
behind this filtering of memory accesses is that when it cannot be
guaranteed that a reference is a hit at level $L$, the worst-case
situation occurs when a cache access to level $L+1$ is performed.
Unfortunately, this assumption is not safe as soon as the degree of
associativity is greater than or equal to two, as
shown on the counterexample depicted in Figure~\ref{fig:limit}.

The figure represents possible streams of memory references on a
system with a L1 2-ways associative cache and a L2 2-ways associative cache, both 
with a LRU replacement policy. The safety problem is observed on
reference $x$, assumed to be performed inside a function. References
$a$, $b$, $c$, and $d$ do not cause any safety problem (they cause misses in
the L1 and L2 both at analysis time and at run-time); they are introduced only
to illustrate the safety problem on reference $x$. Let us assume that:

\debliste
\item $a$ and $c$ map onto the same set as $x$ in the L1 cache and in the L2 cache.
\item $b$ and $d$ map onto the same set as $x$ in the L1 cache and map onto
  a different set than $x$ in the L2 cache.  This frequent case may
  occur because the size of the L1 cache is smaller than the size of
  the L2 cache.
\finliste

The left part of the figure presents the contents of the abstract
cache states at points $p1$, $p2$, $p3$ and $p4$ in the reference stream
(only the sets where reference $x$ is mapped are shown for the sake of
conciseness), as well as the resulting classification. In the
figure, $\{a,x\}$ means that both $a$ and $x$ may be in the cache line.
The right part of the figure presents the concrete cache contents
at the same points when the worst-case execution path (WCEP), which takes the right path in the conditional construct, is followed.

From the classification of reference $x$, the analysis outcome is 2
misses in the L1 cache + 2 hits in the L2 cache. In contrast, executing the worst-case
reference stream results in 1 hit in the L1 cache + 1 miss in the L1 cache + 1 miss
in the L2 cache. Assuming an architecture where a miss is the worst-case 
and $2*Thit_{L2} < Tmiss_{L2}$, the contribution to the WCET of the
cache accesses to $x$ when executing the code is larger than the one
considered in the analysis, which is not safe. This counterexample has
been coded, in order to check that the counter-intuitive behavior of
\cite{mueller97timing} actually occurs in practice.

The safety problem found in \cite{mueller97timing} is due to the combination of severals factors: $(i)$ the reference stream characteristics, $(ii)$ considering uncertain accesses as misses, $(iii)$ considering an access to the next level in such cases.

To further explain the reasons of the safety problem, let us define the \textit{set reuse distance} between two references to the same memory block for a cache level $L$ as the position in the set (equivalent to its  way)  of the memory block when the second reference occurs. If the memory block is not present when the block is referenced for the second time then the set reuse distance is greater than the number of ways. For instance, the set reuse distance of $x$ on Figure \ref{fig:limit} at point $p4$ for Mueller's analysis is $3$ in the L1 cache (greater than the number of L1 ways) and $2$ in the L2 cache (present in the second way). In contrast for the possible concrete cache this value is $3$ (not present in L1 cache) and $3$ (not present in L2 cache).
In \cite{mueller97timing}, uncertain accesses are always propagated to the next cache level
and the analysis may underestimate the set reuse distance. This underestimation 
then results in more hits in the next level in the analysis than in a
worst-case execution. Our approach fixes the problem by enumerating the two possible behaviors of every uncertain access (i.e. considering that the access may occur or not).

%------------------------------------------------------------------------ 
\section{Multi-level set-associative instruction cache WCET analysis}
\label{sec:solution}

After a brief overview of the structure of our multi-level cache
analysis framework (\S{}~\ref{sec:Overview}), we define in this section
the classification of memory accesses (\S{}~\ref{sec:Classif}), and
detail the analysis and prove its termination (\S{}~\ref{sec:Analysis}). 
The use of the cache analysis outputs for WCET computation is presented in \S{}~\ref{sec:WCET}. 

\subsection{Overview}
\label{sec:Overview}

Our static multi-level set-associative instruction cache analysis
is applied to each level of the cache hierarchy separately. The approach
analyses the first cache level (L1 cache) to classify every reference
according to its worst-case cache behavior (\textit{always-hit,
  always-miss,} \textit{first-hit, first-miss} and \textit{not
  classified}, see \S{}~\ref{sec:Classif}). This cache hit/miss
classification ($CHMC$) is not sufficient to know if an access to a
memory block may occur at the next cache level (L2). Thus, a
\textit{cache access classification (CAC)} (\textit{Always},
\textit{Never} and \textit{Uncertain}, see \S{}~\ref{sec:Classif}) is
introduced to capture if it can be guaranteed that the next
cache level will be accessed or not.

\begin{figure}[htbp]
\begin{center}
\psfig{width=0.35\textwidth,figure=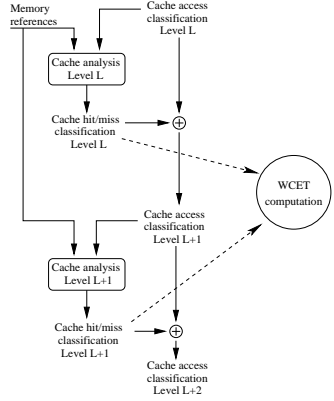}
\end{center}
\vspace*{-0.5cm}
\caption{Multi-level cache analysis framework\label{fig:multianalysis}}
\end{figure}

The combination of the CHMC and the CAC at a given level is used as an input of the
analysis of the next cache level in the memory hierarchy. Once all the cache
levels have been analyzed, the cache classification of each level is used to
estimate the WCET. This framework is illustrated in Figure~\ref{fig:multianalysis}.

\vspace*{-0.5cm}

\subsection{Cache classification}
\label{sec:Classif}

\subsubsection*{Cache hit/miss classification}

Due to the semantic variation of the cache classification between
static cache simulation \cite{mueller00timing} and abstract
interpretation \cite{theiling00fast} approaches, we detail the cache
hit/miss classification (CHMC) used in our analysis, similar to the
one used in \cite{theiling00fast}:

\debliste
\item \textit{always-hit} (AH): the reference is guaranteed to be in cache,
\item \textit{always-miss} (AM): the reference is guaranteed not to be in cache,
\item \textit{first-hit} (FH): the reference is guaranteed to be in cache the
  first time it is accessed, but is not guaranteed afterwards,
\item \textit{first-miss} (FM): the reference is not guaranteed to be in cache
  the first time it is accessed, but is guaranteed afterwards,
\item \textit{not-classified} (NC): the reference is not guaranteed to
  be in cache and is not guaranteed not to be in cache.
\finliste

\subsubsection*{Cache access classification}

In order to know if an access to a memory block may occur at a given
cache level, we introduce a \textit{cache access classification}
(CAC). It is used as an input of the cache analysis of each level to decide if the block has to be considered by the analysis or not. The
cache access category for a reference $r$ at a cache level $L$ is
defined as follows:
\debliste
\item $N$ (Never): the access to $r$ is never performed at cache level $L$,
\item $A$ (Always): the access to $r$ is always performed at cache
  level $L$,
\item $U$ (Uncertain): it cannot be guaranteed that the access to $r$ is
  always performed or is never performed at level $L$.
\finliste

The cache access classification for a reference $r$ at a cache level $L$
depends on the results of the cache analysis of the reference $r$ at the level
$L-1$ (cache hit/miss classification, and cache access classification):

$$
CAC_{r,L}=f(CAC_{r,L-1} , CHMC_{r,L-1} )
$$

The CAC for a reference $r$ at level $L$ is $N$ (never) when the cache
hit/miss classification for $r$ at a previous level is \textit{always-hit}
(i.e. it is guaranteed that accessing $r$ will never require an access
to cache level $L$). On the other side, the CAC for a reference $r$ at
level $L$ is $A$ for the first level of the cache hierarchy, or when
CHMC and CAC at level $L-1$ are respectively 
\textit{always-miss} and $A$ (i.e. it is guaranteed that accessing will always
require an access to cache level $L$).  The CAC for reference $r$ at
level $L$ is $U$ in all the other cases, expressing the uncertainty
that the cache level $L$ is accessed. As detailed in \S{}~\ref{sec:Analysis}, 
the cache analysis for $U$ accesses explores the two cases where $r$ accesses cache level $L$ or not, to identify the worst-case.

Table \ref{tab:classif} shows all the possible cases of cache access
classifications for cache level $L$ depending on the results of the
analysis of level $L-1$ (CACs and CHMCs).

\begin{table}[hp]
\centering
\small
\begin{tabular}{|c||c|c|c|c|c|}
\hline
\backslashbox{$CAC_{r,L-1}$}{$CHMC_{r,L-1}$}  & AM & AH & FH & FM & NC \\
\hline
A & A & N & U & U & U\\
\hline
U & U & N & U & U & U\\
\hline
N & N & N & N & N & N\\
\hline
\end{tabular}
\caption{Cache access classification: level~L}
\label{tab:classif}
\end{table} 

The table contents motivate the need of the cache access
classification. Indeed, in case of an \textit{always-miss} at level
$L-1$, determining if a reference $r$ should be considered at level
$L$ requires more knowledge than the CHMC can provide: if $r$ is always referenced
at level $L-1$ ($CAC_{r,L-1}=A$), it should also be considered at
level $L$; similarly, if it is unsure that $r$ is referenced at level
$L-1$ ($CAC_{r,L-1}=U$), the reference is still unsure at level $L$.

\begin{figure*}[htbp]
\begin{center}
\psfig{width=0.8\textwidth,figure=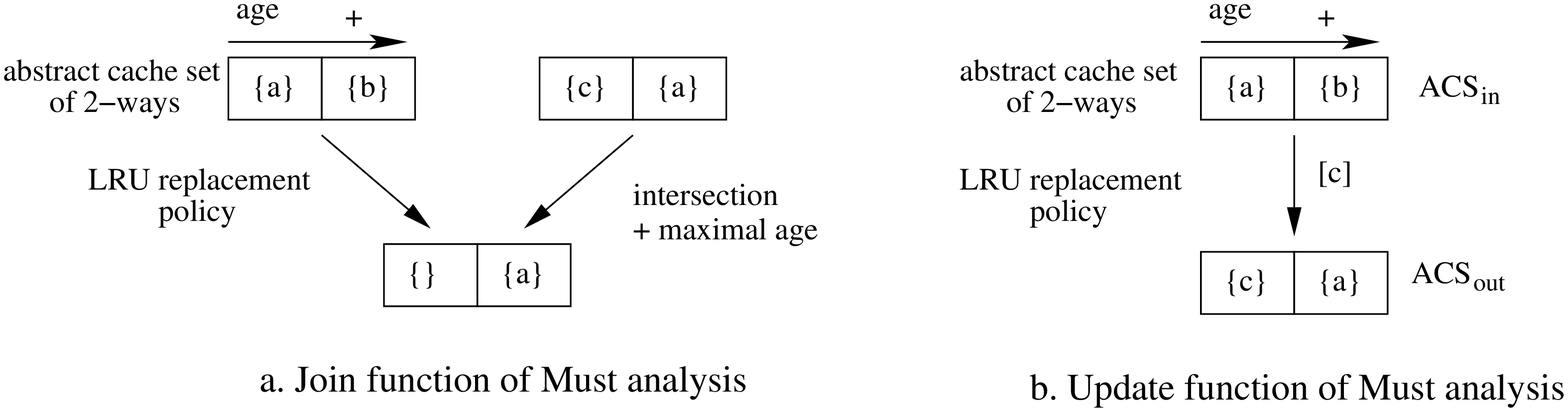}
\end{center}
\vspace*{-0.5cm}
\caption{$Join$ and $Update$ functions for the Must analysis with LRU replacement\label{fig:mustExample}}
\end{figure*}

It also has to be noted that in the case of a $N$ access, the cache hit/miss 
classification can be disregarded because the value will be ignored 
during the WCET computation step for the considered level.

\subsection{Multi-level analysis}
\label{sec:Analysis}

The proposed multi-level analysis  is based on a well known cache analysis method. The analysis
presented in \cite{theiling00fast} is used, due to the theoretical
results of abstract interpretation \cite{CousotCousot77-1,CousotCousot04-WCC}, and the support for multiple replacement policies \cite{theiling00fast,HECK:03} (LRU, Pseudo-LRU, Pseudo-Round-Robin). 
Nevertheless, our analysis can also be
integrated into the static cache simulation method
\cite{mueller00timing}.

The method detailed in \cite{theiling00fast} is based on three
separate fixpoint analyses applied on the program control flow graph:

\debliste
\item a \textit{Must} analysis determines if a memory block is always
  present in the cache at a given point: if so, the block CHMC is
  \textit{always-hit};
\item a \textit{May} analysis determines if a memory block may be in
  the cache at a given point: if not, the block CHMC is
  \textit{always-miss}. Otherwise, if not present at this point in the Must analysis and in the Persistence analysis the block CHMC is \textit{not classified};
\item a \textit{Persistence} analysis determines if a memory block
  will not be evicted after it has been loaded; the CHMC of such
  blocks is \textit{first-miss}.
\finliste

Abstract cache states are computed at every basic block.  Two
functions on the abstract domain, named $\mathit{Update}$, and
$\mathit{Join}$ are defined for each analysis:
\debliste
\item Function $\mathit{Update}$ is called for every memory reference
  on an ACS to compute the new ACS resulting from the memory reference. This function considers both the cache
  replacement policy and the semantics of the analysis.
\item Function $\mathit{Join}$ is used to merge two different
  abstract cache states in the case when a basic block has two predecessors
  in the control flow graph, like for example at the end of a
  conditional construct.
\finliste

Figure \ref{fig:mustExample} gives an example of the $\mathit{Join}$
(\ref{fig:mustExample}.a) and $\mathit{Update}$
(\ref{fig:mustExample}.b) functions for the \textit{Must} analysis 
for a 2-ways set-associative cache with LRU replacement policy. As in this
context sets are independent from each other, only one set is
depicted. A concept of \textit{age} is associated with the cache block of the same set. The smaller the block age the more recent the access to the block.
For the \textit{Must} analysis, a memory block $b$ is stored only once in the ACS, with  its maximum age. It means that its actual age at run-time will always be lower than or equal to its age in the ACS. The $\mathit{Join}$ and $\mathit{Update}$ functions are defined as follows for the \textit{Must} analysis with LRU replacement (see Figure~\ref{fig:mustExample}):

\debliste

\item The $\mathit{Join}$ function applied to two ACS results in an ACS containing
only the references present in the two input ACS and with their \textit{maximal} age. 

\item The $\mathit{Update}$ function performs an access to a memory reference $c$ using an input
abstract cache state $ACS_{in}$ (the abstract cache state before the
memory access) and produces an output abstract cache state
$ACS_{out}$ (the abstract cache state after the memory access). 
 The $Update$ function maps $c$ onto its $ACS_{out}$ set with the younger age and increases the age of the other memory blocks present in the same set in $ACS_{in}$. When the age of a memory block is higher than the number of ways, the memory block is evicted from $ACS_{out}$.

\finliste
 
For the other analyses (\textit{May} and \textit{Persistence}), the approach is similar and the $Join$ function is defined as follows:
 
 \debliste
 \item \textit{May} analysis: union of references present in the ACS and with their \textit{minimal} age;
 \item \textit{Persistence} analysis: union of references present in the ACS and with their \textit{maximal} age.
 \finliste
 
 For more details see~\cite{theiling00fast} and for the other replacement policies see~ \cite{HECK:03}.

\begin{figure}[htbp]
\begin{center}
\psfig{width=0.3\textwidth,figure=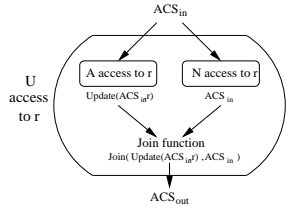}
\end{center}
\vspace*{-0.5cm}
\caption{$Update_m$ function for U access\label{fig:uaccess}}
\end{figure}

Extending \cite{theiling00fast} to multi-level caches does not require
any change in the original analysis framework. Only the base functions
have to be modified to take into account the uncertainty of some
references at a given cache level, expressed by the cache access
classifications (CAC). Function $\mathit{Join}$ needs not be
modified. Function $\mathit{Update}$ (named hereafter $\mathit{Update_m}$ to distinguish our function from the original one) is defined as
follows, depending on the CAC of the currently
analyzed reference $r$:

\begin{itemize}
\item \textbf{A (Always) access.} In the case of an $A$ access the original $\mathit{Update}$ function is used.
  $$ACS_{out}=Update(ACS_{in},r) \mbox{ ; } Update_m  \Leftrightarrow Update$$

\item \textbf{N (Never) access.} In the case of a $N$ access, the analysis does not
  consider this access at the current cache level, so the abstract
  cache state stays unchanged.
  $$ACS_{out}=ACS_{in} \mbox{ ; } Update_m  \Leftrightarrow identity$$

\item \textbf{U (Uncertain) access.} In the case of an $U$ access, the analysis deals with
  the uncertainty of the access by considering the two possible
  alternative sub-cases (see figure~\ref{fig:uaccess} for an
  illustration):
  \debliste
  \item the access is performed. The result is then the same as an $A$ access;
  \item the access is not performed. The result is then the same as a
    $N$ access.
  \finliste

  To obtain the $ACS_{out}$ produced by an $U$ access, we merge this
  two different abstract cache states by the $\mathit{Join}$ function.
\end{itemize}

\vspace*{-0.6cm}

{\small
\begin{eqnarray}
ACS_{out}=Join(Update(ACS_{in},r) , ACS_{in}) \nonumber\\
Update_m(ACS_{in},r) = Join(Update(ACS_{in},r) , ACS_{in}) \nonumber
\end{eqnarray}}

The original functions $Join$ and $Update$ produce a safe hit/miss classification of the memory references. In our case, this validity is kept for the $A$ accesses and is obvious for the $N$ accesses. As for the $U$ accesses, which are the key to ensure safety, the analyses have to keep the semantics of each analysis. For the \textit{Must} and \textit{Persistence} analyses, the $Update_m$ function maintains the maximal age of each memory reference by the original $Join$ function applied to the two ACS (access occurs or not). Similarly, for the \textit{May} analysis, the minimal age is kept by the $Update_m$ function. So the semantic of each analysis is maintained by the $Update_m$ function.

\subsubsection{Termination of the analysis}

It is demonstrated in \cite{theiling00fast} that the domain of abstract cache states is finite and, moreover, that the $\mathit{Join}$ and $\mathit{Update}$
functions are monotonic. So, using ascending chains (every ascending chain is finite) proves the termination of the fixpoint computation.

In our case, the only modification to \cite{theiling00fast} is the
$\mathit{Update}$ function. Thus, to prove the termination of our
analysis we have to prove that the modified function $\mathit{Update_m}$ is monotonic
for each type of cache access.

\textbf{Proof:}~for an $A$ access, $\mathit{Update_m}$ is identical to
$\mathit{Update}$, so it is monotonic. For a $N$ access
$\mathit{Update_m}$ is the identity function, so it is
monotonic. Finally, for an $U$ access, $\mathit{Update_m}$ is a
composition of $\mathit{Update}$ and $\mathit{Join}$. As the composition
of monotonic functions is monotonic, $\mathit{Update_m}$ is then also
monotonic. 
This guarantees the termination of our analysis for each type of cache access and thus for the whole analysis.
\cqfd
It is important to note that our analysis terminates for any monotonic $Update$/$Join$ functions. Thus, all $Update$/$Join$ functions defined in \cite{theiling00fast,HECK:03} to model different replacement policies can be directly reused.

\subsection{WCET computation}
\label{sec:WCET}

The result of the multi-level analysis gives the worst-case access
time of each memory reference to the memory hierarchy. In other words,
this analysis produces the contribution to the WCET of each memory
reference, which can be included in well-known WCET computation methods
\cite{PUSC:97a,84850}.

In the formulae given below, the contribution to the WCET of a NC
reference at level L is the latency of an access to level L+1, which
is safe for architectures without timing anomalies caused by
interactions between caches and pipelines, as defined in
\cite{LUND:99a}. For architectures with such timing anomalies
(e.g. architectures with out-of-order pipelines), more complex methods
such as \cite{LI:06a} have to be used to cope with the complex
interactions between caches and pipelines.

\begin{table*}[hbtp]
\centering
\small
\begin{tabular}{|l|p{8cm}|l|}
\hline
\textbf{Name} & \textbf{Description} & \textbf{Code size} \\
     &             & (bytes)\\
\hline
matmult & Multiplication of two 50x50 integer matrices
& 1200\\

ns   & Search in a multi-dimensional array
& 600\\

bs   & Binary search for the array of 15 integer elements
& 336\\

minver & Inversion of floating point 3x3 matrix
& 4408\\

jfdctint & Integer implementation of the 
forward  DCT (Discrete Cosine Transform)
& 3040\\

adpcm & Adaptive pulse code modulation algorithm
& 7740\\

\hline

task1   & Confidential
& 12711\\

task2   & Confidential
& 12395\\

\hline

\end{tabular}
\caption{Benchmark characteristics}
\label{tab:tasks}
\end{table*}

We define the following notations: constant $Thit_{\ell}$ represents the cost in cycles of a hit at level $\ell$ (accesses to the main memory are always hits), $first$ and $next$ to distinguish the first and the successive execution in loops, the binary variables $first\_present_{\ell}(r)$ and $next\_present_{\ell}(r)$  represent that an access to reference $r$ occurs (1) or not (0) at level $\ell$. Finally, variables $COST\_first(r)$ and $COST\_next(r)$ give the contribution to the WCET of a reference $r$ at a given point in the program, that can be used to compute the WCET. $COST\_first(r)$ and $COST\_next(r)$ are computed as follows:

{\small
$$
COST\_first(r) = \sum_{\ell=1}^{n} Thit_{\ell} * present\_first_{\ell}(r)
$$}

\vspace*{-0.5cm}

{\small
$$
COST\_next(r) = \sum_{\ell=1}^{n} Thit_{\ell} * present\_next_{\ell}(r)
$$}

$first\_present_{\ell}(r)$ and $next\_present_{\ell}(r)$ are computed as follows:

{\small
$$
present\_first_{\ell}=
\left\{
\begin{array}{ll}
1 \mbox{ if } & \ell=1 \\
1 \mbox{ if } & present\_first_{\ell - 1}=1 \\ 
 & \wedge{} \mbox{~~} (CHMC_{\ell - 1}=AM \\
  & \mbox{~~~~~~}\vee CHMC_{\ell - 1}=FM  \\
  & \mbox{~~~~~~}\vee CHMC_{\ell - 1}=NC)  \\
0 & \mbox{ otherwise}  \\
\end{array}
\right.
$$}

\vspace*{-0.5cm}

{\small
$$
present\_next_{\ell}=
\left\{
\begin{array}{ll}
1 \mbox{ if } & \ell=1 \\
1 \mbox{ if } & present\_next_{\ell - 1}=1 \\ 
 & \wedge{} \mbox{~~} (CHMC_{\ell - 1}=AM \\
  & \mbox{~~~~~~}\vee CHMC_{\ell - 1}=FH  \\
   & \mbox{~~~~~~}\vee CHMC_{\ell - 1}=NC)  \\
0 & \mbox{ otherwise}  \\
\end{array}
\right.
$$}

%------------------------------------------------------------------------ 
\section{Experimental results}
\label{sec:perfs}

In this section, we evaluate the tightness of our static multi-level
cache analysis comparatively to the execution in a
worst-case scenario. We also evaluate the extra computation
time caused by the analysis of the cache hierarchy. We first describe the experimental
conditions and then we give and analyze experimental results.

\subsection{Experimental setup}

\paragraph{Cache analysis and WCET estimation.} 

The experiments were conducted on MIPS R2000/R3000 binary code
compiled with gcc 4.1 with flag O0. The
WCETs of tasks are computed by the Heptane\footnote{Heptane is an
  open-source static WCET analysis tool available at {\em
    http://www.irisa.fr/aces/software/software.html\/}.} timing
analyzer~\cite{COLI:01b}, more precisely its Implicit Path Enumeration
Technique (IPET). The fixpoint analysis is an
implementation of the abstract interpretation approach initially
proposed in \cite{theiling00fast}. The \textit{Must}, \textit{May} and
\textit{Persistence} analysis are conducted sequentially on a
two-level cache hierarchy (L1 and L2 caches), both caches implementing
a LRU replacement policy.  The analysis is context sensitive (function are
analyzed in each different calling context).

To separate the effect of the caches from those of the parts of the
processor micro-architecture, WCET estimation only takes into account
the contribution of caches to the WCET as presented in Section~\ref{sec:WCET}. The effects of other
architectural features are not considered. In particular, we do not
take into account timing anomalies caused by interactions between
caches and pipelines, as defined in \cite{LUND:99a}.  The cache
classification \textit{not-classified} is thus assumed to have the same
worst-case behavior as \textit{always-miss} during the WCET
computation in our experiments.

The computation time measurement is realized on an Intel Pentium 4 3.6 GHz with 2 GB of RAM.

\paragraph{Measurement environment.} 

The measure of the cache activities on a worst-case execution scenario
uses the Nachos educational operating
system\footnote{Nachos web site,
  http://www.cs.washington.edu/homes/tom/nachos/}, running on top of a
simulated MIPS processor. We have extended Nachos with a two-level
cache hierarchy with a LRU replacement policy at both levels.

\paragraph{Benchmarks.} The experiments were conducted on five small
benchmarks and two tasks from a larger real application (see Table
\ref{tab:tasks} for the application characteristics). All small benchmarks are benchmarks maintained by
M\"alardalen WCET research
group\footnote{http://www.mrtc.mdh.se/projects/wcet/benchmarks.html}. The
real tasks are part of the case study provided by the automotive
industrial partner of the Mascotte ANR
project\footnote{http://www.projet-mascotte.org/} to the project
partners.

\subsection{Results}

\paragraph{Precision of the multi-level analysis.} 

In order to determine the tightness of the multi-level analysis,
static analysis results are compared with those obtained by executing
the programs in their worse-case scenario.  Due to the difficulty to
identify the input data that results in the worst-case situation in
complex programs, we only use the simplest benchmarks
(\textit{matmult, ns, bs, minver, jfdctint}) to evaluate the precision
of the analysis.

Small L1 and L2 instruction caches are used in this part of the
performance evaluation in order that the code of most of the benchmarks (except
\textit{ns} and \textit{bs}) do not fit into the caches. The L1
cache is 1KB large, 4-ways associative with 32B lines. We use two different L2 caches configurations of 2KB 8-ways associative: one with 64B lines and another one with 32B lines.

\begin{table}[hbtp]
\centering
\small
\begin{tabular}{|c|c||c|c||c|c|}
\hline
 \textbf{Benchmark} & \textbf{Metrics} & \textbf{Static Analysis} & \textbf{Measurement} & \textbf{Static Analysis} & \textbf{Measurement} \\
  & & \textbf{32B - 64B lines} & \textbf{32B - 64B lines} & \textbf{32B - 32B lines} & \textbf{32B - 32B lines} \\
\hline
\textbf{jfdctint} & nb of L1 accesses & 8039 & 8039      & 8039 & 8039 \\
 & nb of L1 misses & 725 & 723          & 725 & 723 \\
 & nb of L2 misses & 54 & 49               & 101 & 96 \\
\hline
& cache contribution to WCET & & & & \\
&  L1+L2, cycles &  20689 &  \multirow{2}*{20169}       & 25389 &  \multirow{2}*{24869} \\
&  L1 only, cycles & 87789 &  & 87789 & \\
\hline
\textbf{bs} & nb of L1 accesses & 196 & 196       & 196 & 196\\
& nb of L1 misses & 16 & 11                                   & 16   &  11\\
& nb of L2 misses & 15 & 6                                     &   16    & 11\\
\hline
& cache contribution to WCET & & & & \\
& L1+L2, cycles &  1856 &   \multirow{2}*{906}       & 1956 &  \multirow{2}*{1406} \\
& L1 only, cycles & 1956 &  & 1956 & \\
\hline
\textbf{minver} & nb of L1 accesses & 4146 & 4146     & 4146 & 4146 \\
& nb of L1 misses & 150 & 140                                         & 150    & 140 \\
& nb of L2 misses & 108 & 71                                            & 150  & 140 \\
\hline
& cache contribution to WCET & & & & \\
&  L1+L2, cycles &  16446 &  \multirow{2}*{12646}        & 20646 &  \multirow{2}*{19546} \\
&  L1 only, cycles & 20646 &  & 20646 & \\
\hline
\textbf{ns} & nb of L1 accesses & 26428 & 26411               & 26428 & 26411\\
& nb of L1 misses & 23 & 13                                                    & 23        & 13\\
& nb of L2 misses & 20 & 7                                                      & 23         & 13\\
\hline
& cache contribution to WCET & & & & \\
&  L1+L2, cycles &  28658 &  \multirow{2}*{27241}        & 28958 &  \multirow{2}*{27841} \\
&  L1 only, cycles & 28958 &  & 28958 & \\
\hline
\textbf{matmult} & nb of L1 accesses & 525894 & 525894        & 525894 & 525894 \\
& nb of L1 misses & 51 & 41                                                            & 51 & 41 \\
& nb of L2 misses & 49 & 19                                                             & 51 & 38 \\
\hline
& cache contribution to WCET & & & & \\
& L1+L2, cycles &  531304 &   \multirow{2}*{528204}       & 531504 &  \multirow{2}*{530104} \\
& L1 only, cycles & 531504 &  & 531504 & \\
\hline
\end{tabular}

\bigskip

\centering
\small
\begin{tabular}{|c|c||c|c|}
\hline
 \textbf{Benchmark} & \textbf{Metrics} & \textbf{Static Analysis}  & \textbf{Static Analysis} \\
 & & \textbf{32B - 64B lines} & \textbf{32B - 32B lines} \\
\hline
\textbf{adpcm} & nb of L1 accesses & 187312  & 187312 \\
 & nb of L1 misses & 2891                                     & 2891 \\
 & nb of L2 misses & 289                                        & 297 \\
\hline
& cache contribution  to WCET & &  \\
&  L1+L2, cycles &   245122                               & 245922 \\
&  L1 only, cycles &   505322                               & 505322 \\
\hline

\textbf{task1} & nb of L1 accesses & 1872522  & 1872522 \\
 & nb of L1 misses & 678                                       & 678 \\
 & nb of L2 misses & 662                                       & 678 \\
\hline
& cache contribution  to WCET & &  \\
&  L1+L2, cycles &   1945502                                & 1947102 \\
&  L1 only, cycles &  1947102                                & 1947102 \\
\hline

\textbf{task2} & nb of L1 accesses & 6783    & 6493 \\
 & nb of L1 misses & 792                                  & 796 \\
 & nb of L2 misses & 718                                  & 796 \\
\hline
& cache contribution  to WCET & &  \\
&  L1+L2, cycles &  86503                                 & 94053 \\
&  L1 only, cycles &  93903                                 & 94053 \\
\hline
\end{tabular}

\caption{Precision of the static multi-level n-ways analysis (4-ways L1 cache, 8-ways L2 cache. Cache sizes of 1KB/2KB in top table, 8KB/64KB in bottom table).}
\label{tab:result}
\end{table}

To evaluate the precision of our approach, the comparison of the hit
ratio at the L2 level between static analysis and measurement is not
appropriate. Indeed, the inherent pessimism of the static cache
analysis at the L1 level introduces some accesses at the L2 level that
never happen at run-time.  Instead, the results are given in
Table~\ref{tab:result} using two classes of metrics:

\debliste

\item The number of references and the number of misses at every level of the memory hierarchy in the worst-case execution scenario (top three lines) to show the behavior of the multi-level cache analysis.

\item The contribution of the memory accesses to the WCET (bottom 2 lines) when considering a cache hierarchy (L1+L2) and when ignoring the L2 cache (L1 only) to demonstrate the usefulness of multi-level analysis. To compute it, we use a L1 hit cost of 1 cycle, a L2 hit cost of 10 cycles and a memory latency of 100 cycles. When considering only one cache level, the memory latency is 110 cycles.

\finliste

Two types of behaviors can be observed:

\debliste
\item The first type of situations is when the number of L1 misses
  computed statically is very close to the measured value (benchmark
  \textit{jfdctint}). In this benchmark, the base
  cache analysis applied to the L1 cache is very tight. As a
  consequence, the reference stream considered during the analysis of
  the L2 cache is very close to the accesses actually performed at
  run-time. Thus, the number of misses in the L2 is also very close to
  the number of L2 misses occuring during execution. In this case, the
  overall difference between static analysis and execution is mainly
  due to the pessimism introduced by considering the cache hierarchy
  (classification as $U$ of every access that cannot be garanteed to
  be or not to be in the L1).

\item The second type of situations occurs when the static cache analysis at L1
  level is slightly less tight. Then, this behavior is also present at the L2 level and it is increased by the introduction of the $U$ accesses.
  In this case, the multi-level analysis is still tight enough. Moreover it turns out that a lot of
  accesses, not detected as hits by the L1 analysis, can be detected as
  hits by the L2 analysis.
  The resulting WCET is thus much smaller
  than if only one level of cache was considered.

  \finliste

\begin{figure*}[htbp]
\begin{center}
\psfig{width=0.33\textwidth,figure=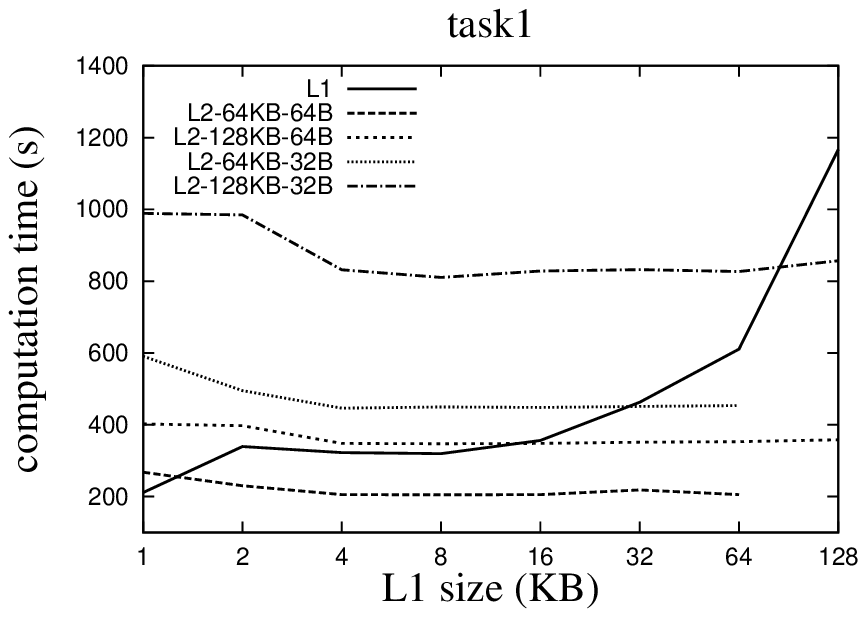}%ColdLoopCommand
\psfig{width=0.33\textwidth,figure=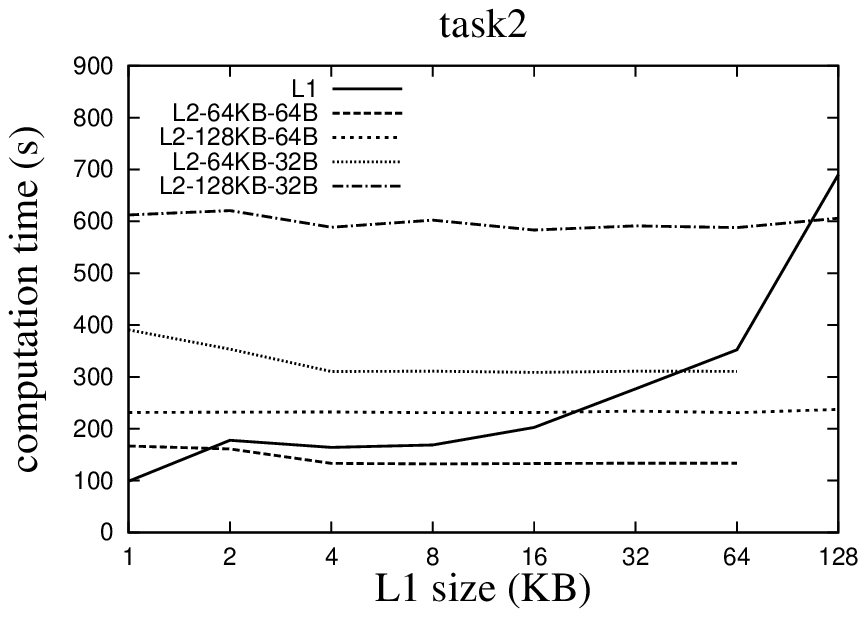}%ClimaetControlManual
\psfig{width=0.33\textwidth,figure=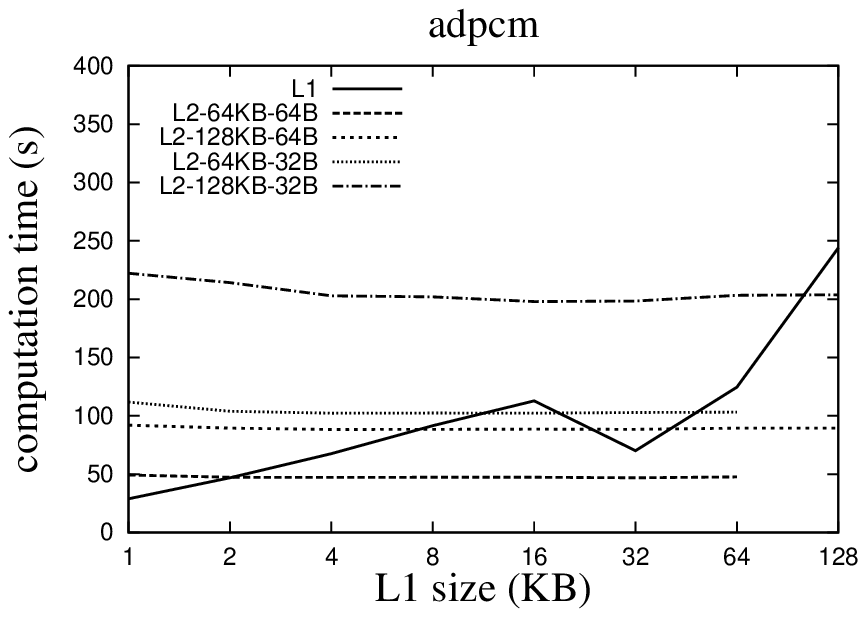}
\end{center}
\vspace*{-0.25cm}
\caption{Computation time with a 64KB and a 128KB L2 cache\label{fig:time}}
\vspace*{-0.25cm}
\end{figure*}

For the largest codes (\textit{adpcm, task1, task2}), only results of
static cache analysis are given (measurements are not realized due to
the difficulties to execute these tasks in their worst-case execution
scenario). Since code size of these three tasks is larger than of the
simple benchmarks, the cache size is now larger and more realistic
than the one considered before.  We use a 8KB large L1 cache and a
64KB large L2 cache with the same cache line sizes and associativity
as before.

We can notice the rather low number of cache hits in the L2 with the
L2 cache with 32B lines. This explains by the size of loops in the
applications as compared to the L1 cache size. In all tasks but {\em
  adpcm}, the code of the loops entirely fits into the L1 cache and
thus there is no reuse once a piece of code gets loaded into the L2
cache. When the cache line size in the L2 cache is larger, the number
of hits increases significantly, due to the spatial locality of
applications.

  In summary, the overall tightness of the multi-level cache analysis
  is strongly dependent on the initial cache analysis of
  \cite{theiling00fast}. In all the cases: $(i)$ the extra pessimism
  caused by our multi-level analysis for the sake of safety
  (introduction of $U$ accesses) is reasonable, $(ii)$ considering the
  cache hierarchy generally results in much lower WCETs comparatively to
  considering only one cache level and an access to main memory for
  each miss.

\paragraph{Computation time evaluation.} 

The analysis time is evaluated on a two-level cache hierarchy, using the
three largest codes (\textit{adpcm}, \textit{task1}, and \textit{task2}) and the same cache structures as before. 
What we wish to evaluate is the extra-cost for analysing the second
level of cache comparatively to a traditionnal cache analysis of only
one level. The extra-analysis time mainly depends on the number of
references considered when analysing the L2 cache, which itself 
depends on the size of the L1 cache (the larger the L1, the higher the
number of references detected as hits in the L1 and thus the lower the
number of references considered in the analysis of the L2). Thus, we
vary the size of the L1 (4-ways and cache lines of 32B) from 1KB to L2
cache size.

Figure \ref{fig:time} details the results for 64 KB (32B and 64B line) and 128 KB (32B and 64B line) L2 caches respectively.
The X axis gives the L1 cache size in KB. The Y axis reports the computation time in seconds.

The shape of the curves are very similar for each used benchmark and
each L2 cache size tested. The computation time for analysing the L1
cache increases with the size because of the inherent dependency of
single-level cache analysis to the cache size. However, the
computation time increase is not always monotonic, like for instance
for benchmark {\em adpcm}. This non-monotonic behavior comes from a
variation of the number of iterations in the fixpoint computation
present in the single-level cache analysis.  In contrast, the analysis
time of the L2 cache decreases when the L1 cache is increased: as the
L1 cache filters more and more memory references, the number of
accesses to the L2 cache considered in the analysis are reduced (more
and more accesses become $N$ access).

The proposed multi-level cache analysis introduced an extra computation cost for $U$ accesses to explore the two possible behavior of uncertain accesses. It can be observed that this extra cost is not visible because it is masked by the filtering of accesses.

When the L2 cache size is 128 KB the slope of the L2 curve is lower than for a 64 KB cache.
This is due to the incompressible time needed for single-level cache analysis of the L2 cache, dependent on the L2 cache size, which masks the filtering effect of the L1 cache.
Nevertheless even in this case the computation time is reasonable.

To conclude, the computation time required for the multi-level set-associative cache (L1 + L2) analysis is significant but  stays reasonable on the case
study application.

%------------------------------------------------------------------------- 
\subsection{Discussion}
\label{sec:discu}

The safety issue of \cite{mueller97timing} is hard to detect on existing codes because of 
$(i)$ the pessimism introduced by the cache
analysis at the first cache level which masks the WCET underestimation caused by the safety issue and $(ii)$ the difficulties to execute tasks in their worst-case condition. We have implemented the
counterexample presented in Section~\ref{sec:limitation} which demonstrates that this phenomenon occurs in practise.

The experiments were undertaken with a LRU replacement policy at
each level of the cache hierarchy. Nevertheless, the modification of the $Update$
function is done at a high level and is independent from any cache
replacement policy.

Finally, experiments were conducted by considering two levels of caches. We
did not present experiments with a L3 cache due to the difficulty
of finding large enough publically available codes. Nevertheless, our method allows the analysis of
a cache hierarchy with more than two levels.

%------------------------------------------------------------------------- 
\section{Conclusion}
\label{sec:conclu}

In this paper we have shown that the previous method to analyze multi-level caches for real-time systems \cite{mueller97timing} is unsafe for set-associative caches. We have proposed a solution to produce safe WCET estimations of set-associative cache hierarchy whatever the degree of associativity and the cache replacement policy.
We have proven the termination of the fixpoint analysis and the experimental results show that this method is precise in many cases, generally tighter than considering only one cache level, and has a reasonable computation time on the case study. In future research we will consider unified caches by using for instance partitioning techniques to separate instruction from data, and we will extend this approach to analyze cache hierarchies of multicore architectures.

%------------------------------------------------------------------------- 
%\nocite{ex1,ex2}
\bibliographystyle{latex8}
\bibliography{latex8}

\newpage
\tableofcontents

\end{document}